\documentclass{iaus}
\usepackage{graphicx}

\title[267~~The active nucleus in NGC 4579] 
{The active nucleus in NGC 4579
}

\author[R. B. Menezes, J. E. Steiner, T. V. Ricci \& A. S. Oliveira]   
{Roberto B. Menezes$^1$, Jo\~ao E. Steiner$^1$, Tiago V. Ricci$^1$
 \and Alexandre S. Oliveira$^2$}

\affiliation{$^1$Instituto de Astronomia Geof\'isica e Ci\^encias Atmosf\'ericas,
Universidade de S\~ao Paulo, Rua do Mat\~ao, 1226, S\~ao Paulo, SP, Brasil \\[\affilskip]
$^2$IP\&D, Universidade do Vale do Para\'iba, Av. Shishima Hifumi, 2911, S\~ao Jos\'e dos Campos,
SP, Brasil \\}

\pubyear{2009}
\volume{267}  
\pagerange{1--1}
\jname{Co-Evolution of Central Black Holes and Galaxies}
\editors{B.M.\ Peterson, R.S.\ Somerville, \& T.\ Storchi-Bergmann, eds.}

\begin{document}

\maketitle


In this work, we present an analysis of a data cube obtained with the IFU/GMOS Gemini North telescope centered on the nuclear region of the LINER galaxy NGC 4579. This galaxy is known to have a type 1 AGN (see Eracleous et al 2002 for a review). The methodology used for the analysis of the data cube was the PCA Tomography (Steiner et al 2009), which consists in applying the statistical tool known as Principal Component Analysis (PCA) to extract information from data cubes. The application of PCA Tomography to the data cube was performed in the following way: first of all, the data cube of this object was transformed into a data matrix, where each row corresponded to a spatial pixel and each column corresponded to a spectral pixel. After that, we applied the PCA in this matrix, obtaining eigenvectors as a function of spectral pixels (which correspond to the original coordinates). These eigenvectors (which have the appearance of spectra) were called eigenspectra. Besides, we calculated the projection of the data (which correspond to the spatial pixels) on each one of the eigenvectors. These images were called tomograms. Figure 1 shows Eigenspectrum E2 and tomogram T2 superposed on T1.

\begin{figure}[h]
\begin{center}
 \includegraphics[width=2.9in, height=1.5
in]{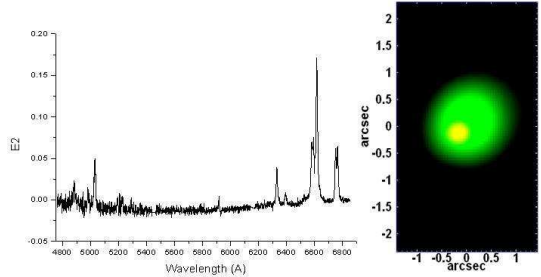} 
 \caption{The eigenspectrum 2 (left), showing the type 1 AGN. Right: Tomogram 2 (AGN) superposed on tomogram 1 (bulge), showing a 0.3¡É displacement. }
   \label{Figure 1}
\end{center}
\end{figure}

Eigenvector E2 shows the presence of a type 1 AGN in this galaxy. The broad wings in H¦Á are clearly seen.  The other lines typically present in LINERS are seen as well. Our analysis also revealed that this AGN is not located at the center of the stellar bulge, being at a distance of 0.3 arcsec from it. This could be due to some differential reddening in the bulge or to a true geometric displacement of unknown origin.

\end{document}